\documentstyle[12pt]{article}
\newcommand{\ka}{\kappa}
\newcommand{\la}{\lambda}

\newcommand{\om}{\omega}
\newcommand{\be}{\begin{equation}}
\newcommand{\ee}{\end{equation}}
\newcommand{\beq}{\begin{eqnarray}}
\newcommand{\eeq}{\end{eqnarray}}

\newcommand{\al}{\alpha}
\newcommand{\Ga}{\Gamma}
\newcommand{\si}{\sigma}

\begin{document}

\title{Some Quantum Aspects of D=3 Space-Time Massive Gravity.}

\vspace{2cm}

\author{ Carlos Pinheiro, \\
Universidade Federal do Esp\'\i rito Santo, \\
Departamento de F\'\i sica,
Vit\'oria-ES, Brazil,
 \\ \\
Gentil O. Pires, \\
Instituto de F\'\i sica, \\ 
Universidade Federal do Rio de Janeiro, \\
Caixa Postal 68528, Cep. 21945-970, \\
Rio de Janeiro-RJ, Brazil
\and  and \\
Fernando A. B. Rabelo de Carvalho, \\
Universidade Federal do Esp\'\i rito Santo, \\
Departamento de F\'\i sica,  
Vit\'oria-ES, Brazil. \\ }

\footnotetext[1]{ fcpnunes@cce.ufes.br }
\footnotetext[2]{ gentil@if.ufrj.br }

\date{ }

\maketitle

\vspace{2cm}

\begin{abstract}
{\it 
We discuss some features of Einstein-Proca gravity in $D=3$ and $4$ space-times.
Our study includes a discussion on the tree-level unitarity and on 
the issue of light deflection in $3D$ gravity in the presence of a
mass term.
}
\end{abstract}

\pagebreak

\vspace{2cm}

\section{Introduction}

It is a well-known fact that the gauge-field approach to pure
Einstein theory of gravitation in D=4 exhibits unitarity at the tree-level, although
renormalizability is not attained. Pure gravity is on-shell finite
only at the 1-loop level [1].

Recently, some effort has been done to analyse Einstein-Chern-Simons
models in D=3 space-time [2], which provides not only
renormalizability, but finiteness and unitarity [3]. Essentially, the
physics of this model consists in giving the graviton a mass
without explicitly or spontaneously breaking gauge symmetry.

Our actual purpose in this paper is to discuss possible consequences
of adding a Proca-like mass term to the Einstein-Chern-Simons  model.
We start our study by analysing the simpler case of Einstein-Proca
gravity in three and four-dimensional space-times (section 2). Then,
in section 3, we concentrate on the Einstein-Chern-Simons-Proca
gravity model in D=3, drawing our attention to the behaviour of the
graviton propagator, in order to infer about unitarity at the
tree-level. The problem of light-ray deflection in a gravitational
field is also contemplated. Finally, in section 4, our concluding
remarks are presented.

\paragraph*{}
\section{The Einstein-Proca Model}

Our starting point is the Einstein-Hilbert action for gravity in D dimensions:
\begin{equation}
\pounds_{{}_{H.E}} = - \frac{1}{2\ka^2} \sqrt{-g} \ R \ ,
\end{equation}
where $R$ is understood to be the scalar curvature. In the weak-field
approach, the metric can be decomposed around the background
flat-space metric as 
\begin{equation}
g^{\mu\nu} (x) = \eta^{\mu\nu} - \kappa h^{\mu\nu} \ .
\end{equation}

So, by replacing (2) into (1), the non-interacting part of the
Lagrangian turns out to be given by the bilinear terms on the
fluctuation $h_{\mu\nu}$ -- field as below:
\begin{equation}
\pounds_{{}_{H.E}}^{\; (2)} = \frac{1}{4}
\partial_{\lambda}h_{\mu\nu}\partial^{\lambda} h^{\mu\nu} -
\frac{1}{4} \partial_{\lambda}h^{\mu}_{\;\; \mu}\partial^{\lambda}
h^{\nu}_{\;\; \nu} + \frac{1}{2} \partial_{\lambda} h^{\lambda}_{\;\; \mu}
\partial^{\mu} h^{\nu}_{\;\; \nu} -
\frac{1}{2} \partial_{\lambda}h^{\lambda}_{\;\; \mu}
\partial_{\nu}h^{\nu\mu} \ .
\end{equation}

If we add by hand a Proca mass term for the graviton,
\begin{equation}
\pounds_{mass} = -\frac{1}{4}\  m^2 (h^{\mu\nu}h_{\mu\nu} -
h^{\mu}_{\;\; \mu} h^{\nu}_{\;\; \nu}) \ ,
\end{equation} 
the complete bilinear Lagrangian can be cast as follows:
\begin{equation}
\pounds = \frac{1}{2} \ h^{\mu\nu} {\cal
O}_{\mu\nu,k\lambda}h^{k\lambda} \ ,
\end{equation}
where ${\cal O}_{\mu\nu,k\lambda}$  can be decomposed on it's Lorentz basis
of spin-operators for
symmetric rank-2 tensors   [4]:
\begin{equation}
{\cal O}_{\mu\nu,k\lambda} = \left\{-\frac{1}{2} (\Box + m^2)
P^{(2)}-\frac{m^2}{2} P^{(1)}_m +  (\Box + m^2) 
P^{(0)}_s + \frac{\sqrt{3}}{2} m^2 \left(P^{(0)}_{sw}
 + P^{(0)}_{ws}\right)\right\}_{\mu\nu,k\lambda} .
\end{equation}           

\noindent 
We stress that, since this theory is not gauge invariant, due to the
Proca-like mass term, we do not need to introduce any gauge-fixing 
term into (6).

The graviton propagator, in configuration space,  is given by:

\begin{equation}
\langle T[h_{\mu\nu}(x) h_{k\lambda}(y)]\rangle =
i{\cal O}^{-1}_{\mu\nu,k\lambda} \delta^D(x-y) \ ,
\end{equation}

\noindent
where the inverse operator ${\cal O}^{-1}$ can be determined with the
help of the multiplicative table of the Barnes and Rivers spin-operators
[4,5]. Basically, one writes the inverse
operator as a linear combination of the spin-operators as below :

\begin{equation}
{\cal O}^{-1}_{\mu\nu,k\lambda} = \left\{X 
P^{(2)} + Y P^{(1)}_m + ZP^{(0)}_s +
WP^{(0)}_w + RP^{(0)}_{sw} + SP^{(0)}_{ws}\right\}_{\mu\nu,k\lambda}
\ .
\end{equation}
Then, with the help of the tensorial identity for  the rank-2 symmetric
tensor,      
\begin{equation}
{\cal O}_{\mu\nu}^{\rho\si}{\cal O}^{-1}_{\rho\si ,k\lambda}
= \left(P^{(2)} + P^{(1)}_m + P^{(0)}_s +
P^{(0)}_w\right)_{\mu\nu,k\lambda} \ ,
\end{equation}
the propagators are uniquely determined. The explicit forms of the operators
appearing above can all be found in Ref. [4,5].

In our case,
\begin{equation}
{\cal O}_{\mu\nu ,k\lambda} =
\left\{AP^{(2)} + B P^{(1)}_m +
C P^{(0)}_s + D P^{(0)}_w  +
EP^{(0)}_{sw} + FP^{(0)}_{ws}\right\} \ ,
\end{equation}
where
\begin{eqnarray}
&& A = -\frac{1}{2} (\Box + m^2) \nonumber \ ; \\
&& B = -\frac{m^2}{2} \nonumber \ ; \\
&& C = (\Box + m^2)  \ ;  \\
&& D = 0 \nonumber \  ;  \\
&& E = \frac{\sqrt{3}}{2} m^2 \nonumber \ ; \\
&&F =  \frac{\sqrt{3}}{2} m^2 \ . \nonumber 
\end{eqnarray}
Then, one  gets the solution:
\begin{equation}
\begin{array}{lll}
X = \displaystyle{-\frac{2}{(\Box + m^2)}} & ; & W =\displaystyle{ -\frac{2(-2\Box + \Box D-2m^2+m^2D)}{m^4(D-1)}}  ;  \\
Y = \displaystyle{-\frac{2}{m^2}} & ; & R = \displaystyle{\frac{2\sqrt{3}}{m^2(D-1)}} ;  \\
Z = \displaystyle{-\frac{2(D-4)}{(D-1)(\Box + m^2)}} & ; & S =\displaystyle{ \frac{2\sqrt{3}}{m^2(D-1)}} .
\end{array}
\end{equation}

\newpage

For $D=3$, the Einstein-Proca propagator reads as below :

\begin{eqnarray}
&& {\cal O}^{-1}_{\mu\nu ,k\lambda} = 
-\frac{1}{(\Box + m^2)} (\eta_{\mu k}\eta_{\nu\lambda}+\eta_{\mu\lambda}\eta_{\nu k}) +
\frac{2}{3(\Box +m^2)} \eta_{\mu\nu}\eta_{k\lambda} + \nonumber \\
&&  + \frac{1}{3} \frac{1}{(\Box + m^2)} \eta_{\mu\nu}\eta_{k\lambda} +
\omega_{\mu\nu}\omega_{k\lambda}
\left[\frac{-2}{\Box + m^2} + \frac{2}{3(\Box + m^2)} + \frac{4}{m^2}\right. +
\nonumber \\
&& + \left. \frac{1}{3} \frac{1}{(\Box +m^2)} -
\frac{(\Box + m^2)}{m^4} - \frac{2}{m^2}\right] +
\eta_{\mu k}\omega_{\nu\lambda} \left[\frac{1}{(\Box +m^2)} -
\frac{1}{m^2}\right] + \nonumber \\
&&  + \eta_{\mu\lambda}\omega_{\nu k} \left[\frac{1}{(\Box +m^2)} -
\frac{1}{m^2}\right] + \omega_{\mu k}\eta_{\nu\lambda} 
\left[\frac{1}{(\Box + m^2)} - \frac{1}{m^2}\right] +  \nonumber \\
&&  + \omega_{\mu\lambda}\eta_{\nu k} 
\left[\frac{1}{(\Box + m^2)} - \frac{1}{m^2}\right] +
\eta_{\mu\nu}\omega_{k\lambda} 
\left[\frac{-2}{3(\Box +m^2)} - \frac{1}{3(\Box + m^2)} +
\frac{1}{m^2}\right] + \nonumber \\
&& + \omega_{\mu\nu}\eta_{k\lambda} \left[\frac{-2}{3(\Box + m^2)} -
\frac{1}{3(\Box + m^2)} + \frac{1}{m^2}\right] \ .
\end{eqnarray}

Proceeding along the same lines, it can be found that the 
propagators for Einstein gravitation [5] in $D=4$, $m=0$; $D=3$, $m=0$;
and for Einstein-Proca [5],
$D=3$, $m\neq 0$ and $D=4$, $m\neq 0$ (Einstein-Proca) are respectively 
given by:
\begin{eqnarray}
&& \langle T[h_{\mu\nu}(-k) h_{k \lambda}(k)]\rangle = 
\frac{i}{k^2} (\eta_{\mu k}\eta_{\nu\lambda} +
\eta_{\mu\lambda}\eta_{\nu k} - \eta_{\mu\nu}\eta{k \lambda}) ; 
 \\
&&   \langle T[h_{\mu\nu}(-k) h_{k\lambda}(k)]\rangle = 
\frac{i}{k^2} (\eta_{\mu k}\eta_{\nu\lambda} +
\eta_{\mu\lambda}\eta_{\nu k} - 2\eta_{\mu\nu}\eta{k \lambda}) ; 
 \\
&& \langle T[h_{\mu\nu}(-k) h_{k \lambda}(k)]\rangle = 
 \frac{i}{(k^2-m^2)} (\eta_{\mu k}\eta_{\nu \lambda} +
\eta_{\mu\lambda}\eta_{\nu k} - \eta_{\mu\nu}\eta_{k\lambda});  \\
&&   \langle T[h_{\mu\nu}(-k) h_{k\lambda}(k)]\rangle =
\frac{i}{(k^2-m^2)} (\eta_{\mu k}\eta_{\nu \lambda} +
\eta_{\mu\lambda}\eta_{\nu k} - \frac{2}{3} \eta_{\mu\nu}\eta_{k\lambda}). 
\end{eqnarray}

Coupling these propagators to conserved external sources (energy-momentum tensors),
$T^{\mu\nu}$, one writes the following current-current amplitudes:
\begin{equation}
A_{m=0} = T^{\mu\nu} (\Delta_{\mu\nu,k \lambda}) T^{'k\lambda{}} =
g^2 T^{\mu\nu} \langle h_{\mu\nu} h_{k\lambda}\rangle_{m=0} \ T^{'k\lambda}
\end{equation}
and
\begin{equation}
A_{m\neq 0} = T^{\mu\nu} 
(\Delta_{\mu\nu,k \lambda}) T^{'k \lambda{}} =
\tilde{g}^2 T^{\mu\nu} \langle h_{\mu\nu} h_{k \lambda}\rangle_{m\neq
0} \ T^{'k \lambda} \ .
\end{equation}
Now taking static sources, where only the $T^{00}$ -- component        
contributes, the explicit forms of the amplitudes for massless 
graviton exchange gives us the  relation below for their effective 
coupling constants,
$g$ and $\tilde{g}$:
\begin{equation}
g^2 = \frac{4}{3} \tilde{g}^2 \ .
\end{equation}

We now wish to consider the problem of photons propagating in the gravitational
field of some static matter distribution. In this situation, the propagator describes
graviton exchanges between the photons and the matter that generates the 
gravitational field.
In the metric formalism for General Relativity, light follows geodesic 
lines. In a
field-theoretical treatment, it is the photon-graviton interaction that replaces
this geometrical fact. Now, in (18), $T'_{k\lambda}$ is the usual energy-momentum tensor
for electromagnetism:
\begin{equation}
T'_{k\lambda} = -F_{k\al}F_{\;\; \lambda}^{\al} +
\frac{1}{4} \eta_{k\lambda} F_{\mu\nu} F^{\mu\nu} \ .
\end{equation}
So, in principle,  the transition amplitudes for the deflection of light induced
by massless and massive gravitons are respectively given by:
\begin{eqnarray}
&& A_{m=0} = \frac{i}{k^2} 2g^2 T^{00}T^{'00} \nonumber , \\
&& \\
&& A_{m\neq 0} = \frac{i}{(k^2-m^2)} \frac{8}{3} \tilde{g}^2 T^{00}T^{'00}
\nonumber .
\end{eqnarray}
 However, since in $D=3$ pure Einstein gravity 
there is
no graviton propagation [5], one does not talk about light-ray 
deflection. As a consequence,
in $3D$-gravity, light deflection may be discussed only if the 
graviton becomes massive, the mass being introduced by a Proca-like or 
a topological Chern-Simons term.

 We conclude this section by studying the 
necessary condition to get tree-level unitarity for the Einstein-Proca gravity
in $D=3$.
To do so, we must take the current-current amplitude and analyse the positivity
condition on the imaginary part of the residues at the pole $k^2=m^2$. 
Otherwise, if positivity is not attained, there will be ghosts or non-dynamical
gravitons in the model.

The transition amplitude is
\begin{equation}
A = \tau^{\mu\nu\ast}{(\vec{\boldmath k})}
\langle T[h_{\mu\nu}(-(\vec{\boldmath k}) h_{k\lambda}
(\vec{\boldmath k})]\rangle \tau^{k\lambda} ({\boldmath k}) \ ;
\end{equation}
  
\noindent  where, after some algebraic manipulation, we obtain

\begin{equation}
A = \frac{i}{(k^2-m^2)} \left[2|\tau^{\mu\nu}
(\vec{k})|^2-|\tau^{\mu}_{\mu} (\vec{k})|^2\right] \ .
\end{equation}

Choosing a suitable basis of vectors in momentum  space,
\begin{equation}
k^{\mu} = (k^0;\vec{k}); \ \tilde{k}^{\mu} =
(k^0;-\vec{k}); \ \varepsilon^{\mu}_i = (0;\vec{\varepsilon}_i);  \
i=1, \cdots ,  D-2 \ ,
\end{equation} 
where
\begin{eqnarray}
&& \tilde{k}^{\mu}k_{\mu} = (k^0)^2 + (\vec{k})^2 > 0\nonumber \\
&& \varepsilon^{\mu}_i \varepsilon_{j\mu} = -\delta_{ij} \\
&& k^{\mu}\varepsilon^i_{\mu} = \tilde{k}^{\mu}\varepsilon^i_{\mu} = 0 \ ,
\nonumber 
\end{eqnarray}
and expanding the tensorial current on this basis, we get :

\begin{equation}
\tau_{\mu\nu} = a_{(k)}k_{\mu}k_{\nu} + b_{(k)} k_{(\mu}\tilde{k}_{\nu )} 
+ c_{i(k)}k_{(\mu}\varepsilon^i_{\nu )} +
d_{(k)}\tilde{k}_{\mu}\tilde{k}_{\nu} +
e_{i(k)}\tilde{k}_{(\mu}\varepsilon^i_{\nu )} +
f_{ij}\varepsilon^i_{(\mu}\varepsilon^j_{\nu )} \ .
\end{equation}
The current conservation, $k^{\mu}\tau_{\mu\nu}=0$, at the pole, 
dictates the number of degrees of freedom
among the coefficients $a,b,c,d, e_{i}$ and $f_{ij}$.

It is shown that if $ \left(|c_{i(k)}|^2+|e_{i(k)}|^2 \right) <0$, the necessary condition for
tree-level unitarity, $Res \ Im \ A>0$ at the pole, is automatically ensured.

\section{The Einstein-Chern-Simons-Proca (E.C.S.P.) Gravity}

\paragraph*{}
Let us consider topological gravity in $(2+1)$ and discuss some of its features
whenever a Proca-like mass term is present [6]. The generalized (E.C.S.P.) Lagrangian reads
as below:
\begin{equation}
\pounds = - \frac{1}{2 \ka^2} \sqrt{-g}R + \frac{1}{\mu}\varepsilon^{\lambda\mu\nu}\Ga^{\rho}_{\;\;\lambda\si}
( \partial_{\mu} \Ga_{\rho \;\;\; \nu}^{\;\;\si} + 
\frac{2}{3} 
\Ga_{\mu \;\;\; \om}^{\;\;\si} \Ga_{\nu \;\;\; \rho}^{\;\;\om} ) -
\frac{1}{2}\  m^2 (h^{\mu\nu}h_{\mu\nu}-\xi 
h^{\mu}_{\;\;\mu}h^{\nu}_{\;\;\nu})\ .
\end{equation}

Notice that the massive Einstein-Proca Lagrangian can be recovered by taking
$\mu\rightarrow \infty$. Also, just like before (E.P.), there is no gauge 
invariance. In  this sense, again, a gauge-fixing term is not adjoined. 
In order to obtain the propagators for this model, we just take the bilinear
terms on the quantum field $h_{\mu\nu}$, in (28); this yields the following
operator to be inverted:
\begin{eqnarray}
{\cal O}_{\mu\nu k\la} &=& -\frac{1}{2} (\Box + m^2) P^{(2)} -
\frac{m^2}{2} P^{(1)} +
\left[\Box - \frac{m^2}{2} (1-3\xi )\right] 
P^{(0)}_s -
\frac{m^2}{2} (1-\xi )P^{(0)}_w \nonumber \\
&&  + \frac{\sqrt{3}}{2} m^2 \left(P^{(0)}_{sw} + P^{(0)}_{ws}\right) +
\frac{1}{2M} (S_1+S_2)\ ,
\end{eqnarray}
where $S_1$ and $S_2$ are new spin-operators coming from the Chern-Simons term [5]. The $M$ parameter is the
topological Chern-Simons mass and is just a redefinition of $\mu\left(M= 
\frac{\mu}{8\ka^2}\right)$. To get the Einstein-Proca operator (6),
we must simply take $\xi =1$ and $M$ going to infinity in (29).

The method to invert (29) is the same as the one followed in the previous
section, and so we just need to obtain the coefficients presented in the linear
combination of the whole set of spin  operators:
\begin{equation}
{\cal O}_{\mu\nu , k\la}^{-1} = X_1 P^{(2)} +
   X_2 P^{(1)} + X_3 P^{(0)}_s +X_4
P^{(0)}_w +
X_5 P^{(0)}_{sw}+ X_6 P^{(0)}_{ws} + X_7 S_1 + X_8S_2 \ .
\end{equation}
It can be found that


\begin{eqnarray}
&&X_1 = \frac{-2M^2(\Box +m^2)}{(\Box^2M^2+2\Box m^2M^2 + m^4M^2+\Box^3)}; \\
&& X_2 = \frac{-2}{(m^2)}; \nonumber \\
&& X_3 = A/B; \nonumber \\
&& X_4 = 2\left[\frac{2m^2\xi + \Box -m^2}{m^2[(\xi -1)\Box -3m^2\xi + m^2}\right]; \nonumber 
\\
&& X_5 = \frac{-2\sqrt{3\xi}}{(\Box \xi -3m^2\xi -\Box +m^2)}; \nonumber \\
&& X_6 = \frac{-2\sqrt{3\xi}}{(\Box \xi -3m^2\xi -\Box +m^2)}; \nonumber \\
&& X_7 = \frac{-2M}{(\Box M^2+2\Box m^2M^2 +m^4M^2+\Box^3)}; \nonumber \\
&& X_8 = \frac{-2M}{(\Box M^2+2\Box m^2M^2 +m^4M^2+\Box^3)}. \nonumber
\end{eqnarray}

\vspace{1cm}

\noindent
where $A=\left[(-3+3\xi )\Box^3 + (-4M^2+4\xi M^2) \Box^2 + (4m^2\xi M^2-6m^2M^2)\Box -2m^4M^2
         \right]
      $ 
and
      $B=\left[(-1+\xi )\Box^4 + (m^2-3m^2\xi -M^2 + \xi M^2) \Box^3 + (-m^2M^2-m^2\xi M^2) \Box^2 +
         \right.
      $ 
\linebreak  
      $ \left. + (m^4M^2 - 5m^4M^2\xi )\Box - 3m^6M^2\xi + m^6 M^2
        \right]
      $.

\vspace{1cm}

Next point, we comment on  the dynamical possibility for the graviton in $(2+1)$ to appear as
the mediator of the
gravitational interaction. As before, one couple the propagators to conserved
external currents $\tau^{\mu\nu}(-\vec{k})$ and $\tau^{k\la} (\vec{k})$ and  observe
that only the $X_1$ and $X_3$ coefficients survive the transversallity of these currents. 
So, in the same way as analysed in (23), the dynamics at tree-level
for the E.C.S.P. theory can be discussed.  We need first to analyse poles 
to  get information on the physicallity of the masses. For simplicity, take
$\xi =1$, where $X_1$ and $X_3$ have the same pole structure:
\begin{equation}
\Box^2M^2 + 2\Box m^2 M^2 + m^4M^2 + \Box^3 = 0
\end{equation}
or, in the momentum space,
\begin{equation}
(k^2)^3-M^2(k^2)^2 + 2m^2M^2(k^2) - m^4M^2 = 0 \ ,
\end{equation}
which is a cubic equation in $k^2$. It is checked  that tachyons and ghosts
are excluded from the spectrum whenever $\left|\frac{M}{m}\right| \rangle \frac{3\sqrt{3}}{2}$.
For example, with $\left|\frac{M}{m}\right| =3$, three massive poles can be obtained:
\begin{eqnarray}
&& k^2_1 = 6,4113m^2\ , \nonumber \\
&& k^2_2 = 0,7737m^2, \\
&& k^2_3 = 1,8154m^2 \nonumber \ .
\end{eqnarray}
Therefore, in principle, there is the possibility of having three massive excitations
mediating the gravitational interaction in $(2+1)$ (E.C.S.P.). 
Also, one can verify  that these three massive gravitons satisfies the condition of
positive-definite residue
of their amplitudes [6] that guarantees  tree-level unitarity.
 Finally, we could say that, by power-counting (just by observing the
$X_1$ and $X_3$ coefficients for $D=3$), the propagator is proportional to
$\frac{1}{k^4}$, which for asymptotic values of $k$, leads to a  renormalizable
theory.

\section{Concluding Remarks}

\paragraph*{}

We concentrated our efforts in this paper to understand some features of $3D$
massive gravity. Massive graviton propagators are derived and conditions
for the absence of ghosts and tachyons are found out. The analysis of propators
is also used to conclude that light-ray deflection in $D=3$ is possible only
in the context of massive gravity. In the general case  of Proca and Chern-Simons
mass terms, the spectrum of propagating gravitons is discussed and 3 massive
excitations are identified in a special case of parameters.

\section*{Acknowledgements}

The authors acknowledge Dr. J.A. Hela\"{y}el-Neto for
discussions and a critical reading of the manuscript. 
C.P. and G.O.P. are partially supported by the Conselho de 
Desenvolvimento Cient\'\i fico e Tecnol\'ogico, CNPq-Brazil.


\end{document}